# Development of A 16:1 serializer for data transmission at 5 Gbps

Datao Gong[a], Suen Hou[b], Zhihua Liang[a], Chonghan Liu[a], Tiankuan Liu[a], Da-Shun Su[b], Ping-Kun Teng[b], Annie C. Xiang[a], Jingbo Ye[a],

[a] Department of Physics, Southern Methodist University, Dallas TX 75275, U.S.A.
[b] Institute of Physics, Academia Sinica, Taipei 11529, Taiwan

dtgong@physics.smu.edu

## Abstract

Radiation tolerant, high speed and low power serializer ASIC is critical for optical link systems in particle physics experiments. Based on a commercial 0.25 μm silicon-on-sapphire CMOS technology, we design a 16:1 serializer with 5 Gbps serial data rate. This ASIC has been submitted for fabrication. The post-layout simulation indicates the deterministic jitter is 54 ps (pk-pk) and random jitter is 3 ps (rms). The power consumption of the serializer is 500 mW. The design details and post layout simulation results are presented in this paper.

## I. Introduction

The large volume data production in the recent high energy physics experiments requires a high speed data transmission ASIC for digital optical link between the on-detector and off-detector electronics systems. The radiation tolerance of the ASIC becomes more critical along with the increasing of the luminosity of the beam in the experiments. There are two serializer chips used in the Large Hadron Collider (LHC) experiments, GOL and G-link [1][2]. The GOL with a serial data rate at 1.6 Gbps is based on a 0.25 μm bulk silicon CMOS technology with radiation hardening layout. With a built-in laser driver, its power consumption is about 400 mW at 1.6 Gbps. The G-link has been identified to be radiation tolerant for the present ATLAS Liquid Argon Calorimeter (LAr) readout system. This chip consumes about 2.0 watts at 1.6 Gbps. The upgrade of LAr readout system from LHC to supper-LHC requires optical data link to provide 100 Gbps data rate, 60 times higher than the present, with same power consumption budget for each front-end board (FEB)[16]. Neither GOL nor G-link can meet the power consumption budget and data rate requirement. The development of a higher speed and lower power serializer is necessary for the LAr upgrade.

A commercial 0.25 μm silicon on sapphire (SoS) CMOS technology has been identified to be suitable for ASIC development in the radiation environment in the particle physics experiments [3]. This technology has a $f_T$ of 90 GHz which is much faster than that of the bulk silicon CMOS with the same feature size [4]. In this paper we present a design of a 16:1 serializer working at 5 Gbps based on this technology with 500 mW power consumption. This serializer can be used as a key component in high speed transmitter for LAr upgrading data optical link.

## II. Design

The serializer includes a 16:1 multiplexer, a PLL based clock generator and a CML driver as shown in figure 1. The multiplexer receives 16 bit LVDS signals and outputs CMOS level serial data at 5 Gbps. The clock generator provides clock signals whose phases are locked to input LVDS clock signal to the multiplexer. The CML driver is used to drive high speed differential signals though transmission lines to radiation tolerant optical laser driver [17]. To achieve good immunity of the single-event effect (SEE), we use large transistor size and static D-flip-flop in the whole design.

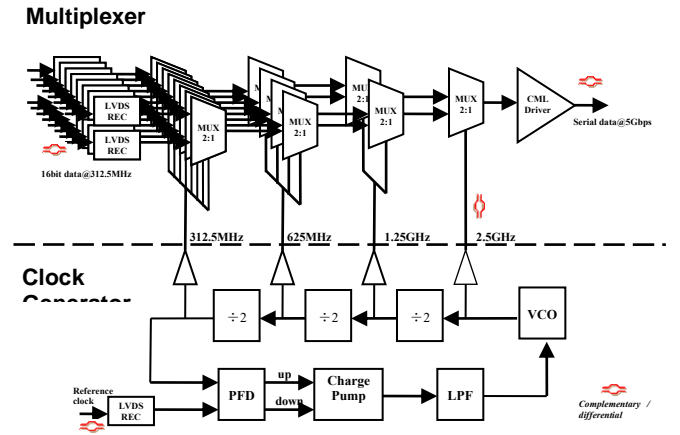

Figure 1: The architecture of the serializer

### A. LVDS receiver

An LVDS receiver is used to convert differential data and reference clock signals to CMOS signals for consequential process. The LVDS receiver is a differential amplifier followed by a differential pair with active load. With 100 mV minimum differential model level requirement, the receiver can work above 400 MHz with common mode level from 0.8 to 1.7 V and consumes about 2.8 mW in the typical corner post-layout simulation.

### B. 16:1 Multiplexer

The 16:1 multiplexer has 4 stage multiplexer units in serial in which the first 3 stages are cascade of same basic CMOS logic 2:1 multiplexer unit and the last stage is a special designed 2:1 multiplexer unit to operate above 2.5 GHz.

The basic 2:1 multiplexer unit is driven by a clock not fast than 2 GHz and converts two input data bits into serial output as shown in figure 2. Two bits data are latched into D-flip-flop at the rising edge of clock signal. One of the two latched data bit is delayed half clock period by a latch to assure the clock signal select data bits in following passive multiplexer with correct timing. The serial output data bit width depends on the duty cycle of clock which requires the clock signal with 50% duty cycle.

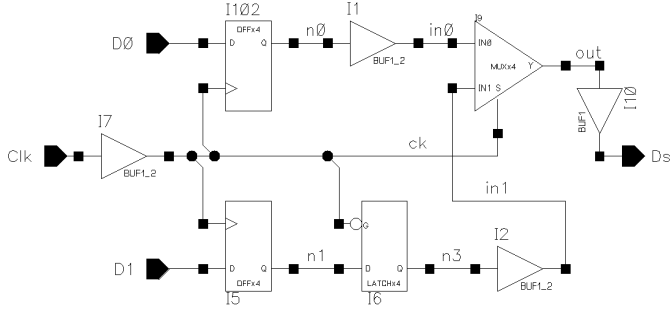

Figure 2: The basic 2:1 multiplexer unit

The static traditional transmission gate D-flip-flop used in the multiplexer unit is fast and has good SEE immunity comparing to other type ones [6][7]. In the regular D-flip-flip, the internal complementary clock signal of the pass gate is generated by inverters and not very symmetric. The asymmetric complementary clock signals significantly increase the delay of the pass-gates switching in the D-flip-flop. This static D flip-flop can not work more than 2 GHz.

A high speed D-flip-flop is required to operate above 3 GHz in the last stage of multiplexer unit and first divider-by-2 circuit following the VCO. A D-flip-flop with symmetrical complementary clock signal inputs meets this requirement as shown in figure 3. We use two identical differential-to-single-ended circuits with cross-couple input from the differential VCO delay stage to generate symmetric complementary clock signals for this unit.

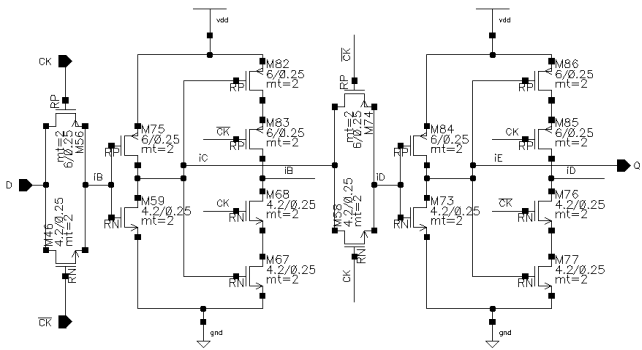

Figure 3: The static D-flip-flop with symmetrical complementary clock signals

### C. Clock generator

The clock generator comprises a PLL and a clock divider. The clock signals distributed from the divider are 312.5 MHz, 625 MHz, 1.25 GHz and 2.5 GHz for four stages of multiplexers respectively. The 2.5 GHz clock signal is complementary signal required by the high speed 2:1 multiplexer unit.

The phase frequency detector (PFD) is dead zone free and maximum operating frequency of this PFD is above 400 MHz. The charge pump requires complementary up and down signals to operate. The asymmetric complementary signals add extra noise on the control node. To minimize the asymmetry of the complementary up and down signals, two inverter arrays are used to generate complementary clock signals as shown in figure 4. After optimization of the transistor sizes, the complementary clock signals match within 5 ps in all process corners [5].

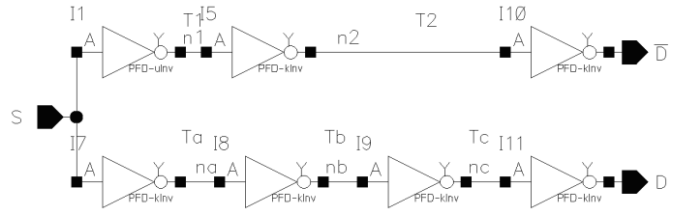

Figure 4: Single-ended to complementary signal converter

A conventional charge pump with active amplifier is implemented. The unitary gain voltage amplifier equalizes the voltage of the mirror node and control voltage node of VCO, which eliminates the charge sharing problem appearing at the instances of switching. Dummy pass-gates are added at the mirror node and control voltage node to reduce the charge injection problem. The current source of the charge pump is programmable from 20 uA to 80 uA to match nonlinear VCO gain. The change pump linear working range is from 0.5 to 2.0 V.

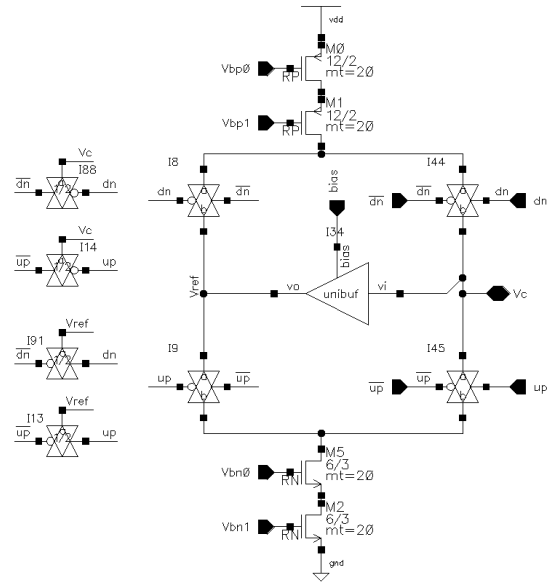

Figure 5: Charge pump with active amplifier

Multiple-pass loop architecture is used in the differential ring oscillator to boost the voltage controlled oscillator (VCO) operating frequency. The extra auxiliary feed forward loop reduces the delay of the stages in a conventional main loop [9][10]. The five stages oscillator is depicted in figure 5. This architecture is also called as look-ahead ring oscillator [11].

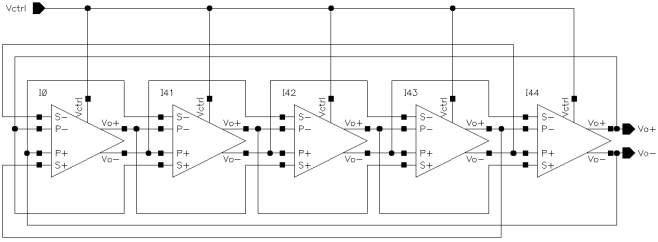

Figure 6: multiple-pass loop 5 stage differential oscillator.

Because of the two loop path structure, two pairs of inputs are needed in the delay stage as depicted in figure 7. Transistors M5 and M6 make the main loop, while M7 and M8 make the secondary loop. Comparing to common differential delay stage, the tail current source is removed, which reduced the phase noise due to the upconversion of the tail transistor low-frequency noise near the oscillation frequency. The oscillating amplitude of this delay stage is rail-to-rail, which also reduce the jitter [13][14].

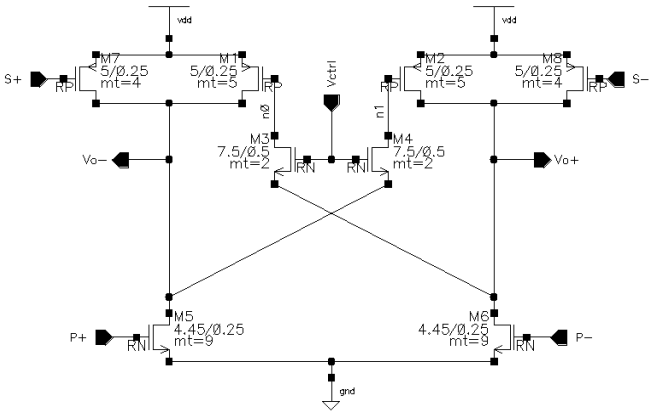

Figure 7: Differential delay stage

As shown in figure 7, transistors M1, M2, M5 and M6 are constructed as a latch. When Vctrl increases, the resistance of M3 and M4 reduces, which increases the positive feedback gain of the latch. The stronger feedback gain makes the latch harder to switch the output nodes. Thus the stage delay increases and the VCO oscillates at a lower frequency when control node voltage increases. The VCO oscillates from 1.5 to 2.75 GHz with the VCO gain varies from 0.4 to 1.1 GHz/V in the charge pump working range. The post-layout simulation indicates that the phase noise is -92 dBc/Hz at 1 MHz offset from the 2.5 GHz carrier frequency.

The PLL low pass filter reduces the low frequency noise for the reference clock, but it is a high pass filter for the VCO generated phase noise. Choosing loop band width is a trade-off among different noise sources. The low pass filter is a bandwidth programmable passive 2$^{nd}$ RC network as shown in figure 8. There is a reset bin to reset the control voltage to Vdd at the initial stage which means the VCO start to oscillate at lowest frequency.

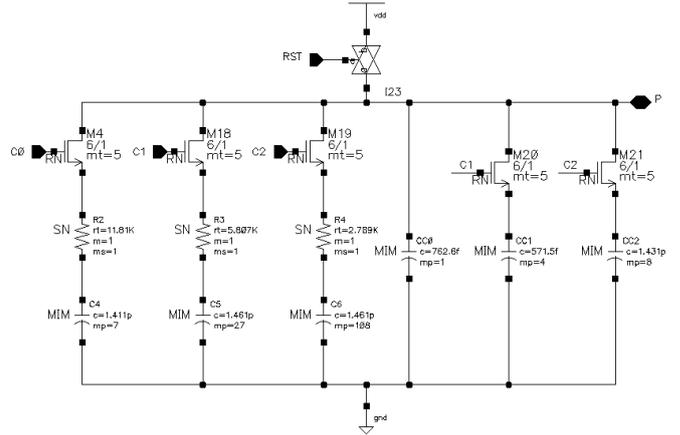

Figure 8: Bandwidth programmable low pass filter

There are 3 control bits C0, C1 and C2 in which only one bit can be set to turn on NMOS transistors switchers and enable the resistors and capacitors in the RC filter. The PLL loop bandwidth and phase margin also depends on the charge pump current. To keep the PLL operating in stable status, its phase margin is larger than 45 degree in all the combination of charge pump current and LPF configurations as shown in table 1.

Table 1: Loop bandwidth in MHz and phase margin in degree with different CP current and the LPF configurations.

| CP current | C0,C1,C2=001 | | C0,C1,C2=010 | | C0,C1,C2=100 | |
|---|---|---|---|---|---|---|
| | BW | margin | BW | margin | BW | margin |
| 20uA | 1.25 | 60 | 2.5 | 60 | 5.0 | 60 |
| 40uA | 2,28 | 56 | 4.6 | 56 | 9.1 | 56 |
| 60uA | 3.14 | 50 | 6.3 | 50 | 12.5 | 50 |
| 80uA | 3.88 | 45 | 7.8 | 45 | 15.5 | 45 |

### D. CML driver

The last stage multiplexer outputs CMOS signals. A followed CML driver is needed to drive the high speed serial data through transmission lines. The CML driver is designed to be 4 stages CML buffer as shown in figure 9 [12]. The following differential stage has twice the current and the transistors are twice in width of those in the previous stage. The last stage amplifier has 20 mA current and 50 Ω output resistance to match the 50 Ω transmission lines outside the chip.

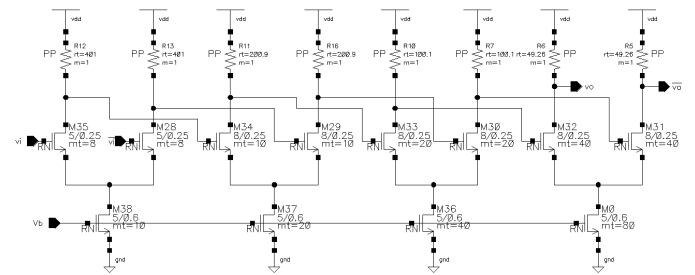

Figure 9: 4 stages CML driver

When the bonding wire with 25 um diameter is 1 mm long, its inductance is about 1nH. The output load assumed in the testing is 1 nH for the bonding wire and 0.2 pF for overall capacitive load that includes the bonding pad and the input capacitance of the optical laser driver module. Resistive load is 50 Ω at the end of an ideal 50 Ω transmission line. The rise and fall times of the output waveform are 44 ps when we test it with 2.5 GHz clock signal. The output signal amplitude at the far-end of the transmission depends on the input frequency as shown in figure 10. As shown in this figure, the CML driver output signal peak to peak amplitude larger than 400 mV at 5 Gbps rate.

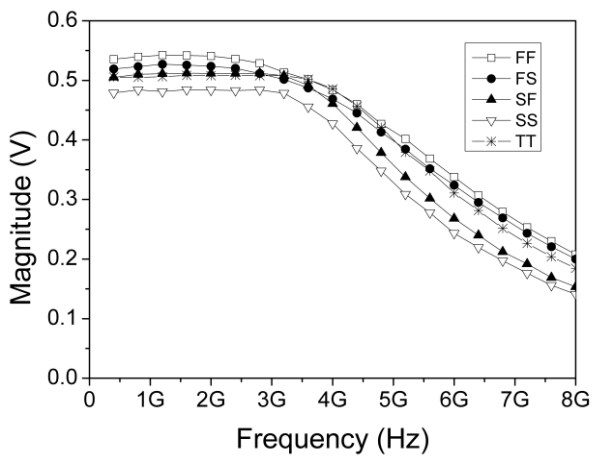

Figure 10: The magnitude of the CML output signal in voltage vs input signal frequency in GHz in different process corners.

The bonding wire length may vary and cause the attached inductance variation. The post-layout simulation manifests that the inductance of bonding wire significantly degenerates the CML driver bandwidth. The bandwidth of the CML driver attached with these bonding wires is about 5.5 GHz. When bonding wire is 5mm, the 3db bandwidth drops to 3.6 GHz. This result suggests us keep the high speed signal bonding wire as short as possible.

## III. PERFORMANCE

The 16:1 serializer is implemented on a 3 mm x 3 mm die and occupies about half of the die area as shown in figure 11. The gray blocks in the plots are decoupling capacitors on the power lines. A high frequency LC-PLL is implemented on the same die for next version of serializer [15]. We separate the multiplexer unit power and ground lines from the noise sensitive PLL circuits to reduce the jitter and noise from the power line. The power consumptions of three main components are shown in table 2. Considering the PLL consumes about 35% power, it is possible to reduce the transmission power by sharing one PLL clock generator with multiple 16:1 multiplexers in the future.

Table 2: Power consumption of serializer components

|  | Power (mW) |
| --- | --- |
| CML Driver | 96 |
| PLL | 173 |
| 16:1 multiplexer | 238 |
| Total | 507 |

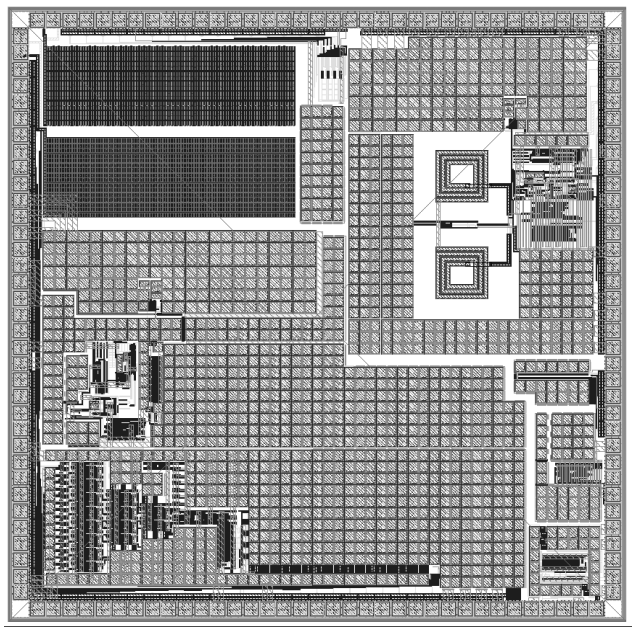

Figure 11: The layout of the serializer, a LC PLL and other test components are included on the same die

In the final post-layout simulation $2^7-1$ PRBS data is feeded through the data bus. The reference clock is 312.5 MHz without jitter. The output load is same as depicted in CML driver testing. To simulate the noise on the power lines the inductance of bonding wires that connect the power and ground lines is considered. Because it is extremely slow to run post-layout simulations with the actual decoupling capacitors which are made of large transistors and metal-insulator-metal device, the on chip decoupling capacitor is simulated with a 0.6 nF ideal capacitor.

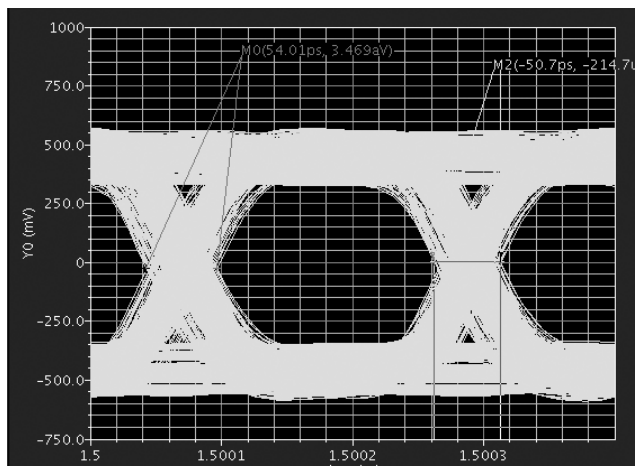

Figure 12: Post-layout simulation with 1nH inductor on each power line and 600pF internal capacitor.

The simulated eye-diagram is shown in figures 12. The transistor noise is not turned on in the simulation thus the jitter quoted in the figure does not include the random jitter. We roughly estimate that the deterministic jitter is 54ps peak-to-peak. Considering the phase noise in post-layout simulation, we estimate that the random jitter from the VCO is less than 3 ps (RMS) in all LPF and charge pump configurations.

## IV. SUMMARY

A 16:1 serializer at 5Gbps is implemented with a 0.25 um SoS CMOS technology. The serializer consumes 500mW power when running at 5Gbps. Its deterministic jitter is estimated to be 54 ps and random jitter is about 3 ps. The design has been submitted for fabrication.

## V. ACKNOWLEDGEMENT

This work is supported by US-ATLAS R&D program for the upgrade of the LHC, and the US Department of Energy grant DE-FG02-04ER41299. We would like to thank Jasoslav Ban at Columbia University, Paulo Moreira at CERN, Fukun Tang at University of Chicago, Mauro Citterio and Valentino Liberali at INFN, Carla Vacchi at University of Pavia, Christine Hu and Quan Sun at CNRS/IN2P3/IPHC, Sachin Junnarkar at Brookhaven National Laboratory, Mitch Newcomer at University of Pennsylvania, Jay Clementson, Yi Kang, John Sung, and Gary Wu at Peregrine Semiconductor Corporation for their invaluable suggestions and comments to help us complete the design work. We also would like to thank Justin Ross at Southern Methodist University for his help in setting up and maintaining the design environment.